# Episodic mass transfer: A trigger for nova outbursts?

Robert Williams
Space Telescope Science Institute
3700 San Martin Drive, Baltimore, MD  21218, USA
e-mail: wms@stsci.edu

**Abstract.** High resolution spectra of postoutburst novae show multiple components of ejected gas that are kinematically distinct. We interpret the observations in terms of episodes of enhanced mass transfer originating from the secondary star that result in the formation of discrete components of circumbinary gas and accretion onto the WD that trigger nova outbursts. In this picture the concordance between absorption line velocities and emission line widths in most novae occurs as a result of the collision of the expanding nova ejecta with a larger mass of surrounding circumbinary gas. One implication of this model is that much of the accreted gas remains on the WD, leading to a secular increase in WD mass over each outburst event. Alternative scenarios to explain nova spectral evolution are possible that do not invoke circumbinary gas and a possible test of different models is proposed.

## 1. Introduction

Observational and theoretical studies of classical novae in recent years have led to a better understanding of the nova phenomenon. Many uncertainties remain but the results of high dispersion spectroscopy in the UV, optical, and IR coupled with detailed numerical modeling of the basic structure, atmospheres, and accretion disks of mass transfer binaries have helped clarify many previously unresolved issues. Further advances in our understanding of cataclysmic variables will certainly come when the Large Synoptic Survey Telescope becomes operational in the coming decade, enabling novae to be characterized before their outbursts, which should provide much information about these objects as their postoutburst studies have in the past.

## 2. The FEROS Survey

The FEROS novae survey (Williams et al. 2008) obtained sequential series of high resolution (R=48,000) optical spectra of 15 novae in 2003-06 with the ESO 2.2m and 1.5m La Silla telescopes and the fiber-fed bench mounted FEROS echelle spectrograph. A major feature of the spectra was the presence of relatively narrow absorption lines from Fe-peak and s-process ions that appear in the great majority of novae and which disappear in the weeks to months following outburst when the initial photospheric P Cygni spectrum converts into the nebular emission spectrum. Analysis of these data led to a geometrical picture of the nova outburst that incorporates as a major component a substantial circumbinary reservoir of transient heavy element absorbing (THEA) gas that appears to pre-exist the nova outburst (Williams & Mason 2010; hereafter 'WM10'). This paper presents a discussion of the interpretation of observed features in the FEROS spectra, including suggestions that circumbinary gas is not prominent, that could be tested with future observations.

## 3. Features of FEROS Spectra

The evolution of novae spectra is dictated by the rapid expansion of an initially optically thick gas. The basic features have been well observed and modeled for individual objects over the years, enabling characteristics of the stars and element abundances to be determined (Warner 1995; Bode & Evans 2008). The FEROS survey focused primarily on one aspect of the spectra: the transient absorbing gas that is evident early in the decline period, and it presented arguments that the THEA gas originates on the secondary star before being ejected into circumbinary orbits, and is *in situ* before the outburst. WM10 speculated on the mechanism that might create the circumbinary reservoir based upon the following facts from the FEROS spectra:
   a) 13 of the 15 novae (85%) observed in the FEROS survey showed (often multiple) THEA systems,
   b) THEA gas is observed to expand outwardly from the novae at velocities of order ~400-1,000 km/s, with most of the systems showing a gradual outward acceleration in velocity in the weeks following maximum light,
   c) in the weeks/months following outburst the THEA systems normally disappear at the time the emission lines appear,
   d) forbidden emission line widths almost always correspond to the same expansion velocities as the THEA absorption lines, and not to the much higher outburst ejecta velocities characteristic of the very broad P Cygni profiles.
   e) column densities derived from THEA line strengths lead to circumbinary gas masses of order $10^{-5}$ $M_\odot$, assuming a solar Fe/H abundance for the secondary star.

   The above characteristics and large circumbinary mass were the basis for the interpretation by WM10 that the emission line spectrum develops when the nova ejecta collide with surrounding circumbinary gas and are decelerated by it due to the latter's higher mass. The frequent delay in the onset of high energy X-rays in novae is a natural consequence of this picture. However, other interpretations of the spectroscopic evolution are possible and an alternative scenario involving the postoutburst wind has been put forward to explain the behavior of the narrow absorption line systems (Shore et al. 2011).

   A basic puzzle of the circumbinary gas picture is: the kinetic energy of expanding THEA gas with the above mass and velocity is $~10^{44}$ erg, which is roughly the total energy released in a normal nova outburst. If the THEA gas is created before the outburst: (i) what is the nature of the creation event?, (ii) what is the source of the required large energy?, and (iii) why are energetic creation events not observed? The most logical answer that satisfies the last two criteria is that the circumbinary gas reservoir is formed by a gradual process in which the kinetic energy of orbital motion of the binary system is converted into heating the outer part of the accretion disk, causing steady ejection of gas from the disk to an outer envelope that orbits the binary system at radii of a few astronomical units.

   Exactly such a process has been proposed and modeled by Artymowicz & Lubow (1996), Dubus, Taam, & Spruit (2002), and by Sytov, Bisikalo, and colleagues at the Moscow Institute of Astronomy for the past decade (Bisikalo 2005; Sytov et al. 2007; 2009). The latter group has performed hydrodynamical calculations of the mass transfer process in close binaries which demonstrate that a bow shock is created along the leading outer edge of the accretion disk by the orbital motion of the WD. As the disk sweeps around it collides with gas whose angular momentum has placed it outside the main body of the disk but inside the WD Roche lobe. The bow shock produces a flow of outer disk gas through the outer L3 Lagrangian point into circumbinary orbits. Angular momentum conservation requires that accretion onto the WD be accompanied by some form of gas excretion or ejection, hence the formation of a circumbinary reservoir confined primarily to the orbital plane occurs inevitably as part of the process that leads to accretion onto the WD.

   The hydrodynamics that create the bow shock could be responsible for the circumbinary THEA gas reservoir that is observed at outburst in most novae. A possible test of this hypothesis is that the process should result in a detectable orbital period change of order $\Delta P/P \sim M_{THEA}/M_{binary} \sim 10^{-4}$ over the epoch of the creation of the reservoir. Such period changes have been studied before in CVs with some success (Schaefer & Patterson 1983) in spite of significant uncertainties caused by effects from the accretion disk (Pringle 1975), and the situation continues to improve (Martin, Livio, & Schaefer 2011).

   Whatever the origin of the THEA gas, its existence in the large majority of postoutburst novae is not in doubt and the energy required to create this component of gas is significant. The absorbing gas is observed to have escape velocities from the binary system when first detected at maximum light, and it is plausible that radiation from the outburst initiates the acceleration from initially bound circumbinary orbits. WM10 calculated that the post-maximum Eddington luminosity normally observed in novae should have sufficient flux to provide for the acceleration of THEA gas.

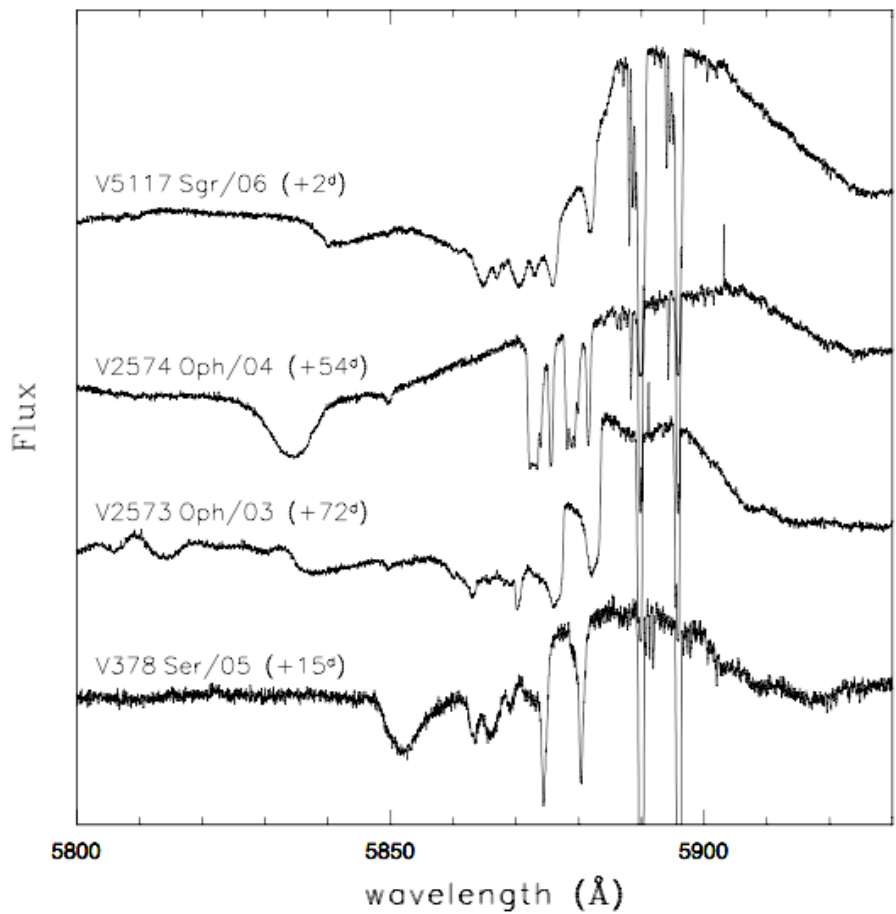

**Figure 1**. Examples of novae with multiple Na I D absorption doublets having discrete radial velocities and different internal velocity distributions. The resolved, narrower doublet systems have velocities substantially less than the expansion velocities of the much broader, unresolved Na I D absorption from the P Cygni profile of the ejecta.

Examination of the FEROS survey spectra shows that of the 85% of novae that were observed to have THEA systems (referred to as the 'principal' spectrum in McLaughlin's nomenclature) a substantial fraction of those showed evidence of multiple absorption systems of different velocities and line strengths, i.e., column densities. Examples of novae with multiple THEA systems are shown in Figure 1. For several of our FEROS survey novae that had strong continua in the blue we succeeded in obtaining data at wavelengths down to the Ca II K line in spite of the poor instrument response caused by the optical fiber attenuation in the blue. We show in Fig. 2 a comparison of the Ca II K and Na I $D_1$ profiles for four novae for which we have data. In most novae the Ca II K absorption exhibits a virtually identical velocity structure to the Na I $D_1$ line. In some instances it is deeper and broader than the corresponding Na I absorption, i.e., the $Ca^+$ is not as confined in velocity as is the neutral $Na^0$, although it always does show a high concentration of absorbing gas at the same velocities as the Na I. The fact that absorption from the ionized species is spread over a larger range of velocities raises the possibility that the narrow THEA systems could be a component of the ejecta rather than due to preoutburst circumbinary gas. We will return to this question.

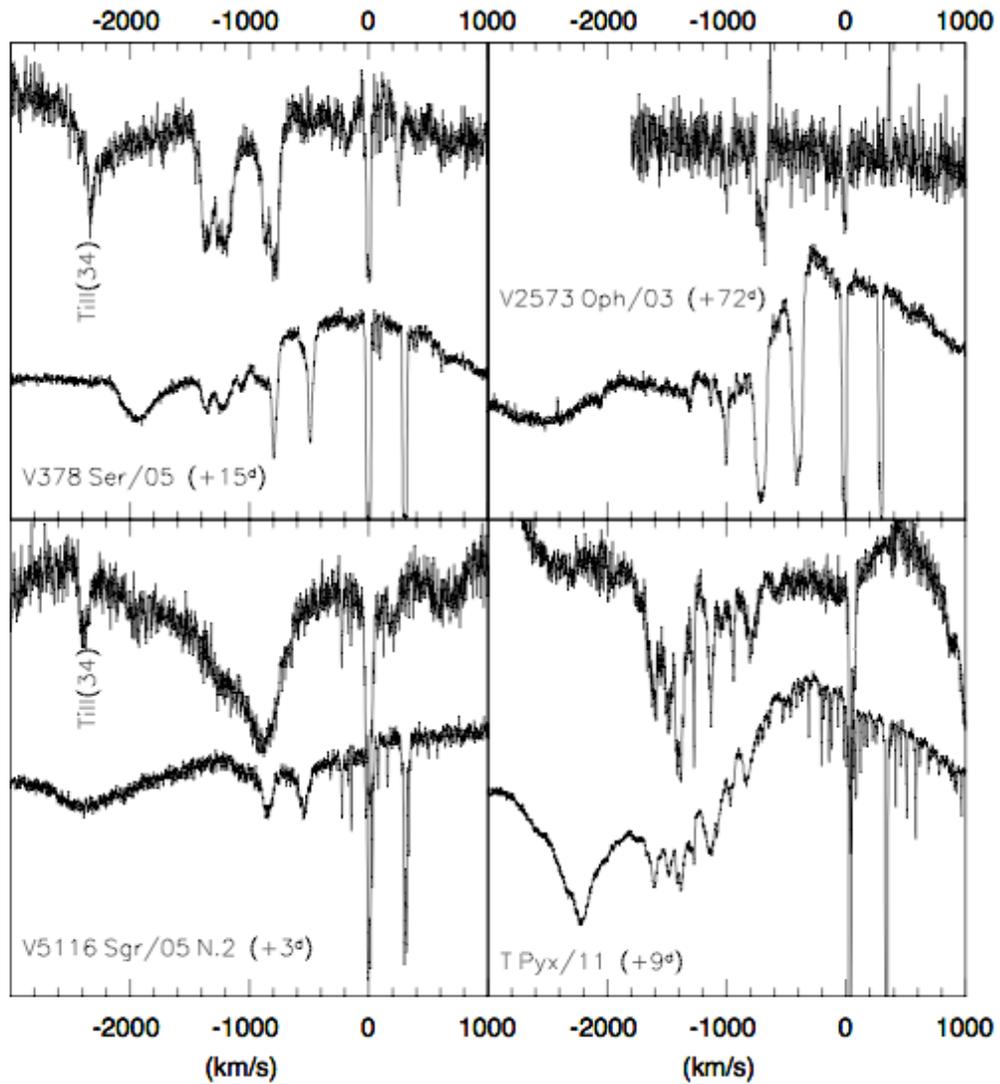

**Figure 2.** Comparative absorption profiles of the Ca II K and Na I D$_1$ THEA lines in four novae. The Ca II and Na I THEA lines generally show a similar velocity structure, in contrast to the profiles of their broader P Cygni absorption components, which are different in the novae.

## 4. Implications for the Nova Outburst

What can be learned about nova outbursts from the fact that their spectra generally show THEA systems? Is circumbinary gas the sequela of processes associated with the outburst or is it a product of the quiescent evolution of the CV system? Might it be a trigger for outbursts? And what is unique about the relatively few novae that show no intervening absorption systems? Assuming that the narrow THEA absorption components are due to gas outside of the ejecta and therefore preexist it, the existence of multiple absorption systems with different velocities is indicative of separate components of gas that could be due to episodes of enhanced mass transfer that are driven by processes in either the secondary star or the accretion disk. The relatively large column densities and masses of ~$10^{-5}$ M$_\odot$ derived for THEA systems, if they are due to circumbinary gas, are difficult to account for by accretion disk events since the masses of disks are normally orders of magnitude less than the above value. Episodes of enhanced mass loss from the secondary star would appear to be the more logical explanation for the existence of kinematically distinct absorption components that precede the outburst.

Conservation of angular momentum considerations, which are confirmed by the numerical calculations of Bisikalo and colleagues (Bisikalo, private communication), require a coupling between the mass of gas accreted onto the WD and the amount that is ejected through the L3 point into the circumbinary reservoir. Thus, when there is evidence for discrete components to a circumbinary envelope in the form of concentrations of density in small velocity intervals, there are likely to be associated accretion episodes that put material onto the WD surface. The magnitude of these mass transfer episodes could be great enough to trigger thermonuclear runaways. Such events could conceivably explain the secondary maxima of novae weeks and months after the primary outburst.

### 4.1. Binary Mass Ratios

Orbital kinetic energy of the WD is required for steady transfer of energy from the bow shock to an outflowing gas stream to occur. Thus, the creation of a circumbinary envelope is strongly favored in systems where the mass ratio $q \equiv M_{sec}/M_{WD} \sim 1$ because, discounting the dynamically unstable $q>1$ regime, comparable stellar masses result in the highest relative orbital velocities of the WD. Values of $q<<1$ leave the WD largely stationary, revolving slowly about the binary center of mass in a small orbit, and only a weak bow shock results (Bisikalo, private communication). Mass transfer certainly takes place through the inner L1 point in this case and should eventually result in an appreciable fraction of the gas ending up on the WD, but with minimal gas ejected from the WD Roche lobe. Presumably, this less commonly observed situation for novae is responsible for outbursts with no THEA systems, as was the case for the recent U Sco outburst, especially when the WD mass is near the Chandrasekhar limit.

The fact that 85% of postoutburst novae are observed to have THEA systems should favor novae in CVs with mass ratios $q \sim 1$ *if* the bow shock mechanism does create the circumbinary reservoir. This prediction offers a test of the hypothesis that THEA systems are formed by enhanced mass transfer that is driven by the orbital motion of the WD. Individual stellar masses of CVs are hard to determine with reliability except for eclipsing systems where the secondary star is detected. The compilation of CVs in the Ritter & Kolb (2003) catalogue lists masses that have been determined by various means that include approximate secondary star masses based upon their spectral classification. Acknowledging the inhomogeneity of the data and the large uncertainty in many of the masses listed in the Ritter & Kolb catalogue Fig. 3 shows histograms of the mass ratio $M_{WD}/M_{sec}$ distribution for novae, nova-like variables, and dwarf novae where individual mass determinations have been made.

The distribution of mass ratios for dwarf novae (DNe), the most common of the CVs, is seen from Fig. 3 to be broad, with two peaks. The group of DNe with $M_{WD}/M_{sec}>4$ consists almost entirely of systems with orbital periods below the 'period gap', i.e., with P< 2 hr. It is seen in Fig. 3 that nova systems loosely occupy a region with ½<q<1 that is more concentrated around values q~1 than the other classes of CVs. Novae below the period gap are rare, which is usually ascribed to the fact that mass transfer in this regime is driven by gravitational radiation and thus occurs at such a low rate that the interval between outbursts is extremely long. Given the large uncertainty in binary mass determinations a great deal of significance cannot be attached to Fig. 3, but it does indicate that few classical novae have q<½.

### 4.2. Secular Evolution of White Dwarf Mass

THEA absorption lines have not been reported previously in quiescent novae systems or nova-like variables although it must be acknowledged that relatively little high resolution spectroscopy of quiescent CVs has been done because of their faintness between outbursts. The presence or absence of THEA systems is important to ascertain because it indicates whether circumbinary gas is a relatively brief phenomenon in the outburst cycle that occurs only around the time of outburst and is therefore ostensibly associated with it. If one accepts that a circumbinary gas reservoir is formed by periods of enhanced mass transfer the absence of THEA systems in quiescent CVs then suggests that such episodes may act as triggers for the nova outburst.

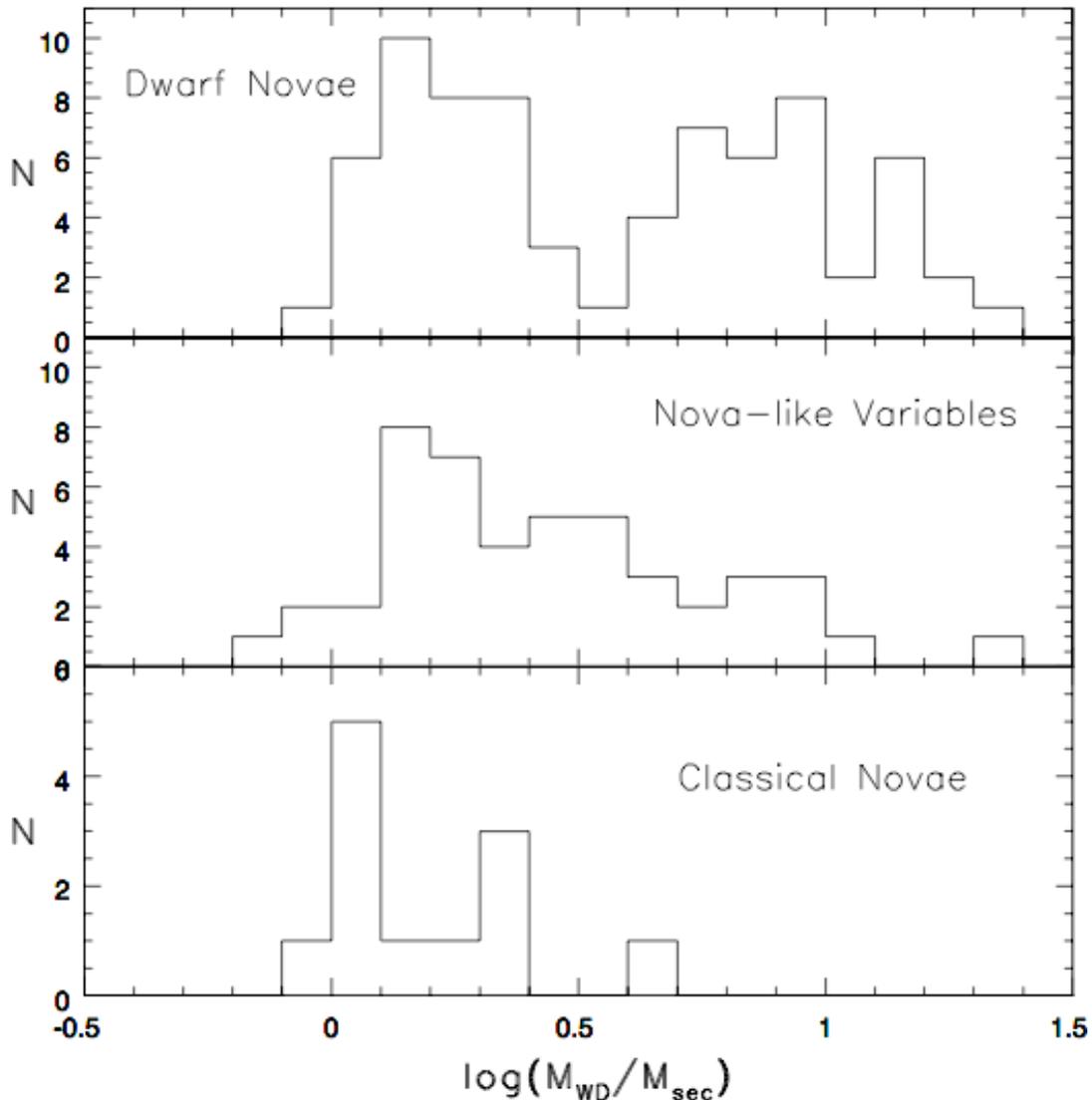

**Figure 3.** Histograms showing the distribution of mass ratios, $M_{WD}/M_{sec}$, for the different classes of CVs. Data are taken from the Ritter & Kolb (2003) catalogue. The admittedly very uncertain data are suggestive that novae may have a higher fraction of objects with mass ratios $q>1/2$ than other CVs, consistent with the expectation that outbursts may be advantaged in CV systems having WDs with higher orbital velocities.

According to the circumbinary scenario for THEA systems an important deduction follows concerning the mass evolution of the WD. Both hydrodynamical calculations and general conservation of angular momentum considerations indicate that the mass of gas accreted onto the WD should be comparable to the amount that is ejected through the L3 point, i.e., the mass of circumbinary gas should be comparable to the mass accreted. WM10 argue that the masses of THEA systems observed in the FEROS survey must be an order of magnitude greater than the masses of the outburst ejecta because of their deceleration of the ejecta upon collision. Since the mass accreted should be comparable to the mass of the circumbinary reservoir, we deduce for the circumbinary scenario that the mass ejected in outbursts is significantly less than the mass accreted. If the colliding shells model of novae is valid, our FEROS observations indicate that the majority of WDs in novae systems increase in mass across each outburst---a conclusion that is contrary to most theoretical outburst calculations.

### 4.3. Origin of THEA Gas: Pre- or Postoutburst?

The interpretation of the THEA systems in novae as ejected circumbinary gas that originates in the secondary star, that forms before and apart from the outburst, and that has a substantial mass exceeding the ejecta mass is consistent with, although not uniquely required by, the observations. The arguments in favor of a preoutburst origin center on (1) the evidence that absorbing gas with small internal velocity dispersions lies outside the faster moving outburst ejecta, and (2) the absence of observed events near outburst that could be associated with ejection of a substantial amount of mass. These attributes occur naturally if there is a steady ejection of gas in quiescent systems that draws upon the orbital energy of the binary.

This circumbinary scenario is not without some difficulties, however. The FEROS survey found circumbinary gas in 85% of the novae studied, indicating that it is distributed along lines of sight with all orientations. Yet hydodynamical studies to date have shown that gas ejected in quiescent systems should be confined largely to the orbital plane (Artymowicz & Lubow 1996; Dubus, Taam, & Spruit 2002; Bisikalo 2005; & Sytov et al. 2007, 2009). One could argue that the sudden injection of outburst energy serves to disperse the gas out of the binary plane, and although this could be true there are as of yet no calculations that demonstrate that the planar gas should end up being distributed over $4\pi$ steradians.

An alternative explanation for the existence of preoutburst gas, directly associated with the outburst TNR but preceding the main ejecta, has been given by Prialnik & Livio (1985) in a study of episodic accretion events that trigger an outburst. They found that episodic accretion can result in a 'bounce' in the accreted gas that ejects of order $\sim 10^{-8}$ $M_\odot$ of infalling gas. An increase in luminosity of more than an order of magnitude occurs that lasts of order one day before the system returns to its quiescent state. Such episodes, if they occur in the weeks before the ultimate outburst TNR could explain the presence of THEA systems if this mechanism can be shown to eject sufficient gas at high velocities, as is observed, with very small internal velocity dispersion. This could obviate the need for a circumbinary reservoir and it might be able to account for the strong relationship between the THEA expansion velocities and the subsequent emission line widths. It might also explain the presence of the Fe-peak and s-process elements in the absorbing gas.

Whether created steadily in quiescence between nova outbursts or as part of the outburst the separate THEA systems with different velocities do argue for multiple formation events. The discrete kinematics of the ejected gas suggest it to be formed episodically. The presence of THEA systems near the time of outburst, coupled with the apparent lack of such absorption in quiescent novae, which needs to be confirmed by observing more quiescent novae and nova-like variables at high resolution, strongly suggests a causal relationship between nova outbursts and episodic mass transfer, i.e., mass transfer episodes emanating from the secondary star may trigger many nova outbursts. In fact, we hypothesize that novae which show THEA systems have outbursts that have been triggered by episodic mass transfer events, whereas those without THEA systems are the result of a TNR initiated by steady mass transfer through the inner L1 Lagrangian point.

### 4.4. Brightness Variations as Discriminator of THEA Origins

The two extremes and most likely scenarios for the origin of THEA systems are either (1) a preoutburst external circumbinary reservoir, or (2) outflowing high-density globules associated with the postoutburst ejecta or wind. The THEA absorption systems are observed against the continuum of the expanding photosphere and depend upon its size in different ways that might be used to discriminate between the two scenarios. The fact that the THEA Na I D doublet intensity ratios show saturation but do not have zero intensity at line centers indicates a typical covering factor of order 25% for the THEA gas. A typical line of sight emanating from an initial small, point-like photosphere is therefore only ~25% likely to intercept an absorbing cloud. However, as the photosphere expands and approaches the size of the putative circumbinary gas ensemble of clouds the likelihood increases that any cloud---and, in fact, multiple clouds---will be seen against the large photosphere. Thus, detection of *circumbinary* THEA systems are favored by a large(r) photosphere.

The situation is the opposite for THEA clouds immersed in the outflowing ejecta. Since they cannot be observed deeper into the ejecta than the τ=1 surface, a larger photosphere disfavors their detection because it obscures all matter deeper within the optically thick ejecta/wind. A smaller photosphere will reveal more absorbing clouds within the flow.

Postoutburst novae experience brightness fluctuations that are likely due to variations in the size and temperature of the photosphere, which in turn are determined largely by the properties of the outflowing gas. Since THEA absorption systems are observed to appear and disappear over short time intervals it would be instructive to determine how the appearance and disappearance of THEA absorption lines is correlated with brightness variations brought about by changes in the size of the photosphere. Good data exist for some novae and therefore tests of the nature and origin of THEA systems should be possible.